\documentclass[12pt,preprint]{aastex}

\usepackage[]{natbib}

\slugcomment{{\it The Astrophysical Journal}}

\shorttitle{}
\shortauthors{Reimer, A.}

\begin{document}

\title{The redshift-dependence of gamma-ray absorption in the environments of strong-line AGN}

\author{A. Reimer}

\altaffiltext{}{W.W. Hansen Experimental Physics Laboratory \& Kavli Institute for particle astrophysics \& cosmology, Stanford University,
452 Lomita Mall, Stanford, CA 94305, USA; afr@stanford.edu}

\begin{abstract}
The case of $\gamma$-ray absorption due to photon-photon pair production of jet photons in the external photon environment like accretion disk and broad-line region radiation field of $\gamma$-ray loud active galactic nuclei (AGN) that exhibit strong emission lines is considered. I demonstrate that this ''local opacity'', if detected, will almost unavoidably be redshift-dependent in the sub-TeV range.
This introduces non-negligible biases, and complicates approaches for studying the evolution of the extragalactic background light with contemporary GeV instruments like e.g. the Gamma-ray Large Area Space Telescope (GLAST), etc., where the $\gamma$-ray horizon is probed by means of statistical analysis
of absorption features (e.g. Fazio-Stecker relation, etc.) in AGN spectra at various redshifts. It particularly applies to strong-line quasars where external photon fields are potentially involved in $\gamma$-ray production.

\end{abstract}


\section{Introduction}
Following the unification scheme the central nucleus of an active galaxy (AGN) consists of a black hole (BH), an accretion disk, line-emitting clouds, a dust torus, and emanates prominent jets when classified as radio-loud. The properties of radio-loud AGN viewed at a small angle to the line-of-sight are in general agreement with the common blazar properties 
\citep[e.g.,][]{Urry95,Padovani07}. Their broadband emission covers the complete electromagnetic band, from the radio up to the $\gamma$-ray band, in some cases even reaching TeV-energies, and is widely dominated by beamed non-thermal emission from a relativistic jet.
The blazar class subdivides into BL Lac objects and flat-spectrum radio quasars (FSRQs).
The difference dividing both subclasses is generally considered in the detection of strong emission lines in the case of FSRQs, while in BL Lac objects the equivalent width of emission lines is depressed, or lines are absent at all. 
The physical reason is thought to lie predominantly in the weak accretion disk radiation field of BL Lac objects, whereas BH in the nuclei of FSRQs accrete with high rates leading to luminous accretion disk photon fields.
The present work deals with $\gamma$-ray loud AGN observed during an epoche of a bright accretion disk and accompagnied
with the apprearance of strong emission lines. For simplicity I will refer to them as ''quasars'' in the following, although some radio-loud AGN classified conventionally as BL Lac objects may occasionally fall into this category as well 
\cite[e.g.,][]{Falomo94,Sbarufatti06}, and vice versa.
 
Gamma-ray production mechanisms, leptonic as well as hadronic ones, in AGN jets often involve either the external radiation fields
associated with the immediate AGN environment or internal jet photons. For example, contributions from interactions in the accretion disk or broad-line region (BLR) radiation fields are often mandatory to explain the overall $\gamma$-ray spectral energy distribution (SED) from FSRQs \citep[e.g.,][]{Ghisellini96,Boettcher00}, while the
SED of low-luminosity BL Lac objects is often fitted with either ''Synchrotron-Self-Compton'' models \citep[e.g.,][]{Ghisellini85} 
or their 
hadronic equivalent, the ''Synchrotron-Proton blazar'' models \citep[e.g.,][]{Mannheim93,Muecke00a,Aharonian00,Muecke03}. The former scenario places the $\gamma$-ray emission region rather close to the BLR clouds. 
This has immediate consequences: If external radiation fields in quasar
environments play a non-negligible role for $\gamma$-ray production, then, at the same time
they are also significant for the quasi-resonant process of photon-photon pair production with its
peak cross section possessing a comparable value to the Thomson cross section.
It should be noted though, that the position
of the high-energy emission region is still a matter of debate, with locations proposed to be also far away from the BLR 
\citep[e.g.,][]{Lindfors05,Sokolov05}. If this is the case, external Compton emission \citep[e.g.,][]{Dermer93} or pair cascade radiation initiated by ultra-high energy cosmic ray interactions on external photon fields
\citep[e.g.,][]{Protheroe97b,Atoyan03} 
 in
those sources does not provide an appreciable contribution to the observed high-energy emission.

The present work concerns specifically opacity features in the $>10$ GeV regime by photon absorption through e$^+$-e$^-$ pair production in radiation fields in the vicinity of the BH, but external to the jets in $\gamma$-ray loud quasars. Past works on this subject \citep[e.g.,][]{Protheroe97a,Donea03,Becker95} indicate the importance of this process for constraining AGN properties like the location of the $\gamma$-ray region above the disk, disk radiation fields, torus temperature, etc. In contrast to these works, I will be focusing on the evolution of external radiation fields in quasar environments, and its consequences
for the resulting opacity features in the $\gamma$-ray band covered by current and near future instruments.
This is primarily motivated by the anticipated studies of the evolution of the extragalactic background light (EBL) through the detection of absorption features in a large sample of high-redshift sources using the Large Area Telescope (LAT) onboard GLAST.
Dedicated methods have been developed here to probe the evolution of the EBL via detecting the horizon of
$\gamma$-rays emitted from extragalactic sources like AGN and GRBs while propagating through the EBL to Earth 
\citep[e.g.,][]{Chen04,Kneiske04}. They involve either the determination of the ratio of absorbed to unabsorbed flux
versus redshift $z$, or the detection of the e-folding cutoff energy $E(\tau_{\gamma\gamma}=1)$ versus redshift 
\cite[''Fazio-Stecker relation'';][]{Fazio70} in a large number of sources at various redshifts in order to disentangle intrinsic blazar features from 
absorption caused ones during propagation in the EBL. The common underlying reasoning for this procedure is that the observation of any redshift-dependent attenuation in $\gamma$-ray AGN can only be attributed to absorption in the EBL, and no other sources of redshift-dependent opacity exist in those sources. Here I will demonstrate that
opacity due to photon absorption from $\gamma\gamma$-pair production caused in external radiation fields within 
the AGN system (in the following called ''local absorption'', to be distinguished from ``self-absorption'' in the internal jet radiation fields) will most likely result in optical depth values that 
increase with the source redshift, and 
coincidentally mimic redshift-dependent EBL-caused absorption. 

The outline of this paper is organized as follows: Sect.~2 and Sect.~3 describe the considered external target photon fields
(accretion disk and BLR radiation field) for photon-photon pair production, and the corresponding optical depth calculation, 
respectively.
In Sect.~5 I will apply various models of supermassive BH growth and accretion rate evolution, that are described
in Sect.~4, to the $\gamma$-ray attenuation calculations. The results with particular emphasis on the possibility of a
redshift-dependence of the local opacity are presented in Sect.~5. The paper closes with conclusions and a discussion 
in Sect.~6.


\section{Characterization of external radiation fields in quasars}

The most relevant target radiation fields in AGN environments
for photon absorption in the LAT energy range, $\sim$0.02-300 GeV\footnote{\url{http://www-glast.slac.stanford.edu/software/IS/glast\_lat\_performance.htm}}, are the optical/UV bands of the accretion disk photon field and the radiation field of the BLR. These will be considered in the following.
I assume the radiation fields to be located azimuthally symmetric with respect to the jet axis, and to radiate
persistently during $\gamma$-ray emission\footnote{Optical continuum variations that are generally attributed to accretion disk radiation in AGN are typically observed on time scales of months to years \cite[e.g.,][]{Peterson93}. This is significantly 
longer than typical variability 
time scales in the $\gamma$-ray domain, and thus allows the disk emission to be approximated as a constant photon field
for the purpose of the present work.}.

\subsection{Quasar accretion disk radiation fields}

The accretion disk spectrum in FSRQs is assumed to follow the cool, optically thick blackbody solution of Shakura \& Sunyaev
 (1973) with a given accretion rate $\dot M_{\rm acc}$, suitable for AGN that show strong emission lines (``strong-line AGN''). 
The differential photon density $n(\epsilon,\Omega)$ into a solid angle $d\Omega=2\pi d\mu$, $\mu=\cos{\xi}$ reads
\begin{equation}
\frac{dn(\epsilon,\Omega)}{d\Omega}=\frac{dn}{2\pi d\mu}=\frac{\epsilon^2}{2\mu c^3}\left(\frac{m_e c^2}{h}\right)^3 
\left[\exp\left(\frac{\epsilon}{\theta(R)}\right)-1\right]^{-1}
\end{equation}
with $R=l\sqrt{\mu^{-2}-1}$, $l$ the distance of the emission region above the BH, $\epsilon$ the photon energy, and
\begin{equation}
\theta(R)=\frac{k_{\rm B}T(R)}{m_e c^2}\simeq 1.44 \left(\frac{M_{\rm BH}}{M_\sun}\right)^{-1/2} \left(\frac{\dot M_{\rm acc}}{M_\sun \rm{yr}^{-1}}\right)^{1/4} \left(\frac{R}{R_g}\right)^{-3/4} \left(1-\sqrt{\frac{R_i}{R}}\right)^{1/4}
\end{equation}
with $R_i=6GM_{\rm BH}/c^2$ for a Schwarzschild metric, $R_g=GM_{\rm BH}/c^2$ and $k_{\rm B}$ the Boltzmann constant. 
For the opacity calculations the disk is considered
to extend from 6 to $10^5 R_g$.
Typical accretion rates for strong quasars approach the Eddington accretion rate $M_{\rm edd}$,
$\dot M_{\rm acc}=0.1-1M_{\rm edd}$, respectively the Eddington luminosity $L_{\rm edd}$.

\subsection{The BLR radiation field in quasars}

The accretion disk is considered as the photo-ionizing source of the BLR material, and emission lines are produced through recombination. The BLR geometry is approximated as a spherical shell with radius $r$ filled with clouds, extending from $r_{\rm in}$ to $r_{\rm out}$.
In this picture the accretion disk as located at $r=0$. In the following the shell size is fixed to $r_{\rm in}=0.01$pc to $r_{\rm out}=0.4$pc, which are typical values for quasars \citep[e.g.,][]{Kaspi04}. A geometrically extended shell is supported from the observations by the AGN Watch campaigns\footnote{\url{http://www.astronomy.ohio-state.edu/~agnwatch/}}.
The opacity calculations require spectral as well as spatial knowledge of the emissivity of the BLR radiation field.
Information about the cloud sizes and their spatial distribution can be deduced in general from reverberation mapping.
Although most details on BLR physics and geometry have been derived by the study of radio-quiet AGN, 
no major differences
in the broad-line flux between radio-quiet and -loud sources have been found so far 
\cite[e.g.,][]{Corbin92,Wills93}.
For the present calculations 
the number density $n_{\rm cl}\propto r^\alpha$ and cross-section $\sigma_{\rm cl}\propto r^\beta$ of the clouds are assumed to follow a power law with exponents $\alpha=-1.5$ and $\beta=0.6$ as derived by Kaspi \& Netzer (1999). 
A fraction $d\tau_{BLR}=n_{\rm cl}\sigma_{\rm cl} dr$ of the central source luminosity $L_{\rm disk}$ is re-processed into line radiation such that the total BLR luminosity, assumed to be optically thin, is $L_{\rm BLR}=\tau_{\rm BLR}L_{\rm disk}$, and these lines are assumed to radiate isotropically. Observational support for a very narrow range of $\tau_{\rm BLR}$, i.e. the ratio of emitted to ionizing continuum photons, is provided by the
observation of a linear correlation between Balmer line and optical disk luminosity in AGN over
several orders of magnitude
\cite[e.g.,][]{Yee80,Shuder81}. This is compatible with photoionization/recombination theory which expects
the H-line brightness to be correlated with the ionizing continuum flux as the line luminosity is driven by
this ionizing disk continuum.
A statistical blazar study of Celotti et al. (1997) implies $\tau_{\rm BLR}\simeq 0.01$, which is used in the following. Changing the here fixed parameters (e.g. $\tau_{\rm BLR}$, $r_{\rm in}$, $r_{\rm out}$, etc.) will alter
the absolute values of the $\gamma$-ray attenuation, however, any redshift-dependence remains 
unaffected.

The calculation of the $\gamma$-ray opacity follows the procedure outlined in Donea \& Protheroe (2003) for the geometrically thick shell case, but uses a more refined BLR line spectrum here. The ''average'' BLR spectrum of Francis et al. (1991)
together with the H$_\alpha$ line strength reported by Gaskell et al. (1981) sums up to 35 lines, with H$_\alpha$ and 
Ly$_\alpha$ being the strongest lines. For the present work the BLR spectrum is approximated
as a two-component spectrum, $n(\epsilon)\propto \delta(\epsilon-\epsilon_{H\alpha}) + \delta(\epsilon-\epsilon_{Ly\alpha})$, 
with the total luminosity of all lines at $>4000~\AA$ ($\sim 32\%$) to be radiated at the strongest optical line (H$_\alpha$) wavelength, and the total line luminosity at $<4000~\AA$ ($\sim 68\%$) emitted at the strongest UV-line, Ly$_\alpha$. 
A refined treatment of the BLR spectrum is straight forward, but will not alter the results on the redshift-dependence
of the local opacity, nor add qualitatively new insights to the subject of the present work.


\section{GeV-photon absorption in quasar radiation fields}

The calculation of the optical depth 
\begin{eqnarray}
d\tau_{\gamma\gamma}(E,l) & = & dl \int_{\epsilon_0}^\infty d\epsilon \int_{\mu_{\rm min}}^{\mu_{\rm max}} d\mu (1-\mu)
\sigma_{\gamma\gamma}(s) 2\pi \frac{dn(\epsilon,l,\mu)}{d\Omega}=\\
& = & \frac{dl}{4E^2}\int_{\epsilon_0}^\infty d\epsilon \epsilon^{-2}\int_{s_{\rm min}}^{s_{\rm max}} \int_{s_{\rm min}}^{s_{\rm max}} s \sigma_{\gamma\gamma}(s) \frac{dn(\epsilon,l,\mu)}{d\mu}
\end{eqnarray}
with 
$E$ the primary $\gamma$-ray photon energy, $\mu=\cos{\xi}$,
$s=2E\epsilon(1-\mu)$, 
$\epsilon_0=\min(\epsilon_{\rm thr},\epsilon_{\rm min})$, 
$\epsilon_{\rm thr}=(2 m_e c^2)^2/(2E(1-\mu_{\rm min}))$, $\epsilon_{\rm min}=s_{\rm min}/(2E(1-\mu_{\rm min}))$, 
$s_{\rm min} = 2E\epsilon(1-\mu_{\rm max})$ and
$s_{\rm max} = 2E\epsilon(1-\mu_{\rm min})$
takes into account the full angle-dependent cross section \citep[e.g.,][]{Gould67,Jauch76}. 
$\mu_{\rm min}$, $\mu_{\rm max}$ are determined by the respective geometry of the target photon field.
The total cross section maximizes at $x=(1-y^{-1})^{1/2}\approx 0.7$, where $y=1/2E\epsilon(1-\mu)/(m_e^2 c^4)>1$ ($\xi$=photon interaction angle) is the threshold condition of the pair production process.
The very prominent peak of the cross section near threshold reaches roughly $\sigma_{\gamma\gamma,\rm max}\approx 0.26\sigma_T$, where $\sigma_T$ is the Thomson cross section.
The narrowness of the pair production cross section forces over half the interactions to occur in a small target photon 
energy interval $\Delta\epsilon\approx(4/3\pm 2/3)\epsilon^*$ centered on $\epsilon^*\approx 0.8(E/{\rm TeV})^{-1}$eV for a 
smooth broadband spectrum \cite{Aha-book}.

\begin{figure}[t]
\resizebox{\hsize}{!}{\includegraphics{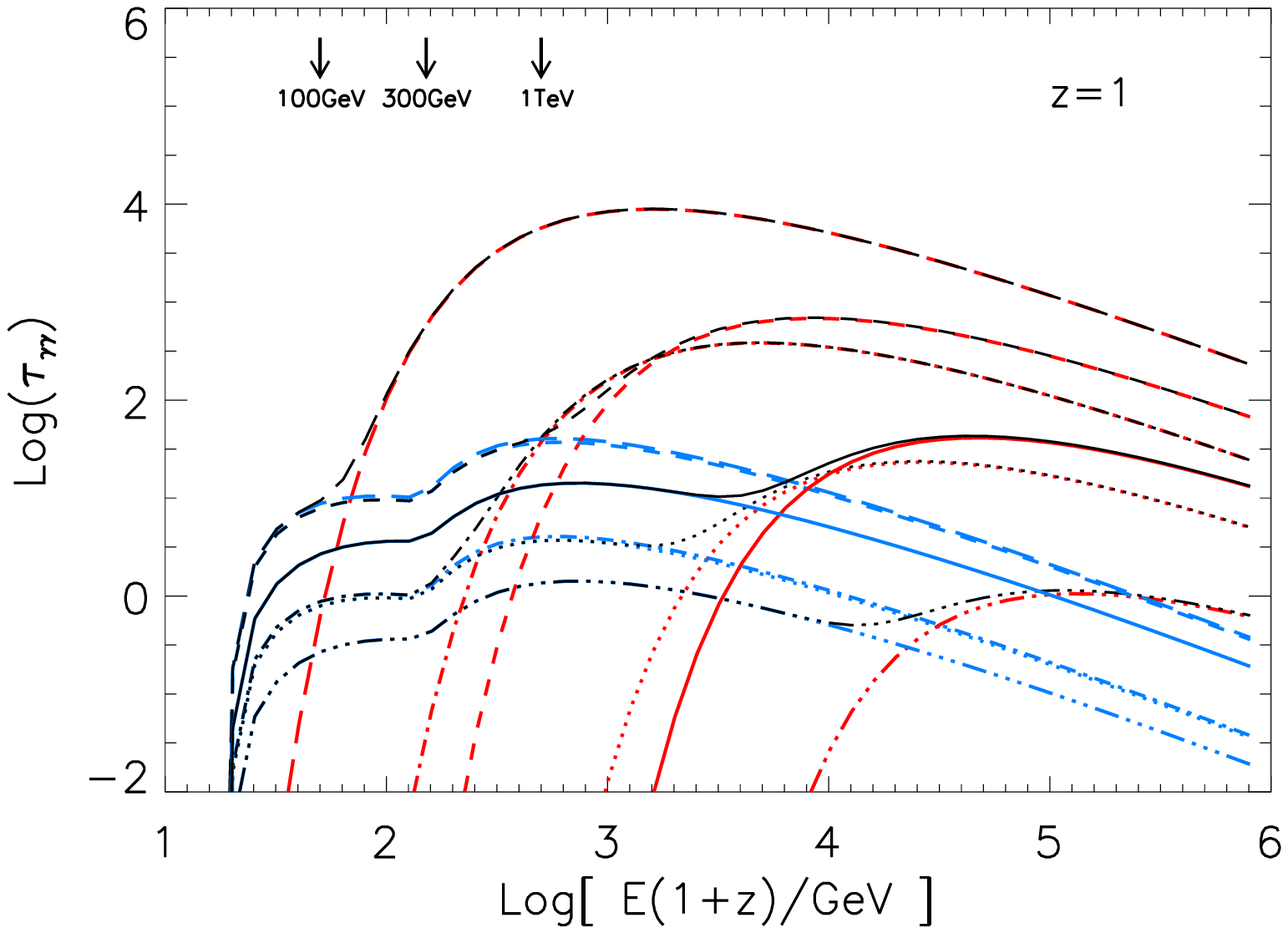}}
\caption{Optical depths from $l=l_0\ldots \infty$ for jet $\gamma$-rays interacting with accretion disk photons (red/pale grey curves) and BLR photons (blue/dark grey curves) with accretion rates taken from Netzer \& Trakhtenbrot (2007) for type-I AGN at redshift $z=1$. The black curves correspond to the sum of both, opacity in the accretion disk and BLR radiation field. Parameters: Black hole mass $M_{\rm BH}=10^8M_\sun (L_{\rm disk}=0.2L_{\rm edd}$, $\dot M_{\rm acc}=0.5M_\sun$yr$^{-1}$) with $l_0=0.1$pc (dashed-triple dotted), $0.01$pc (dotted), $0.001$pc (dashed-dotted) and $M_{\rm BH}=10^9M_\sun (L_{\rm disk}=0.2L_{\rm edd}$, $\dot M_{\rm acc}=5.3M_\sun$ yr$^{-1}$) with $l_0=0.1$pc (solid line), $0.01$pc (dashed) and $0.001$pc 
(long dashed). Energies in the observer frame are indicated by an arrow.}
\label{fig1}
\end{figure}

Fig.~\ref{fig1} shows the resulting opacity from different distances $l_0$ of the $\gamma$-ray production region above the black hole to 
$l\rightarrow\infty$, in the accretion disk (red curves) and BLR radiation field (blue curves) for typical quasar accretion rates 
and BH masses. The two ``bumps'' in the blue opacity curves are the result of absorption in the H$_\alpha$ and Ly$_\alpha$ lines of the BLR,
and smooth when a detailed multi($>30$)-line spectrum is used.
In typical quasar environments the strength of local absorption is strongly dependent on the location of the 
emission region with respect to the target photon field. If the $\gamma$-ray emission region is located not well beyond the BLR, which is 
mandatory for $\gamma$-ray production that involve external photon fields, local $\gamma$-ray absorption features in FSRQ spectra have to be expected at $E(1+z)\geq$ several tens of GeV.

\section{Black hole evolution and accretion rates}

The key step of this work is the application of cosmological black hole and quasar evolution to the expected pair production opacity of $\gamma$-ray photons in AGN, with direct implications for studies of the evolution of the EBL.

With the availability of large AGN data archives enormous advances in BH demographics were made. 
The cosmic evolution of the BH mass accretion rate has recently been studied by Netzer \& Trakhtenbrot (2007) on the 
basis of $\sim 10^4$ SDSS type-I 
radio-loud and radio-quiet AGN in a large redshift range ($z\leq 0.75$). Significantly higher accretion rates at larger
redshifts were derived with an Eddington ratio of the accretion luminosity $L_{\rm acc}/L_{\rm edd} \propto (1+z)^{\delta(M_{\rm BH},z)}$ with $\delta(M_{\rm BH},z)\simeq 6-9$.

The intriguing similarity observed of the time history of star formation (SF) and BH accretion rate density \citep[e.g.,][]{Marconi04} as well as established relations between BH mass and some bulge properties of the host galaxy lead to the widely accepted picture of a joint evolution of QSOs/BHs and their host galaxies \citep[see also][]{Barger01,Marconi04}. 
\footnote{Since the evolution of the EBL is strongly influenced by the cosmic SF history, one may in turn even speculate about an indirect imprint of the BH accretion history into the evolution of the EBL.}
Moreover, the observed BH mass function is consistent with that inferred from quasar luminosities 
for simple assumptions of the accretion efficiency \citep[e.g.,][]{Soltan82,Cavaliere88,Marconi04}. The evolution of their luminosity functions at both, soft and hard X-rays, shows the number density of fainter AGN to peak at lower redshifts than that of the brighter ones \citep[e.g.,][]{Ueda03,Hasinger05,Barger05}. This evidence of ''downsizing'' directly leads to the picture of an
''anti-hierarchical'' BH growth (i.e. high-mass BHs grow faster and low-mass BHs grow preferably at lower redshift), and has meanwhile been confirmed by multiple studies: 
an analysis based on the fundamental plane of accreting BHs \citep{Merloni04};
the phenomenological approach of Marconi et al. (2004) for determining the evolution of the BH mass function using observational constraints from the local BH mass function, the evolving X-ray luminosity functions and energetics from the X-ray background; determinations of low-luminosity high-redshift quasar luminosity functions by the GOODS collaboration \citep[e.g.,][]{Cristiani04}; the COMBO-17 survey at higher luminosities \citep{Wolf03}, just to name a few.
The combination of these observational findings with the theoretically advocated hierarchical clustering paradigm 
(based on cold dark matter) led to the
awareness of feedback processes 
working during the process 
of BH mass growth in AGN systems \citep[see e.g.][]{Granato04,Lapi06,Fontanot06}.

While the parameter space is still large, the advocated evolution of BH growth and thus accretion rates has severe implications for the $\gamma$-ray quasar population where local $\gamma$-ray absorption in accretion disk and BLR radiation fields is potentially important. 

\begin{figure}[t]
\resizebox{\hsize}{!}{\includegraphics{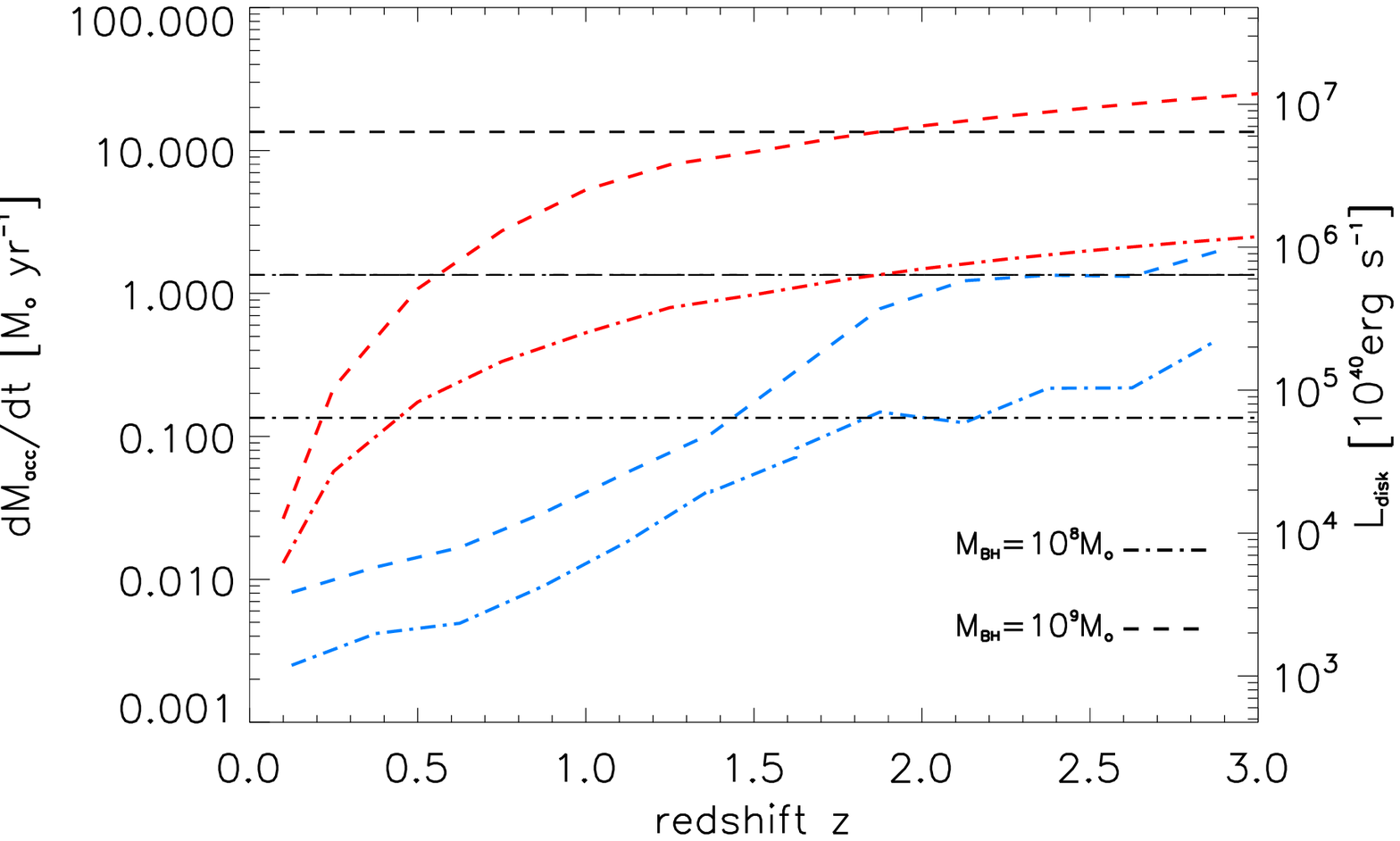}}
\caption{Evolution of BH mass accretion rates. Red/dark grey curves: From the Netzer \& Trakhtenbrot (2007) analysis for $z\leq0.75$, complemented with a mild evolution at large redshifts \citep{Lapi06}. Blue/pale grey curves: Marconi et al. (2004) model. An accretion radiation efficiency of 0.1 was used to convert the average BH growth history into evolution of accretion rates. Black straight lines: no evolution with $\dot M_{\rm acc}=0.1-1M_{\rm edd}$. Dashed curves are for M$_{\rm BH}=10^8$M$_\sun$, dashed-dotted lines for M$_{\rm BH}=10^9$M$_\sun$.}
\label{fig2}
\end{figure}

In the following I use three exemplary models for the evolution of the cosmic BH accretion rate (Fig.~\ref{fig2}): a) 
the Netzer \& Trakhtenbrot (2007) analysis, complemented by the Lapi et al. (2006) model for redshifts $>1$ where only modest evolution is proposed there, b) the ''anti-hierarchical'' BH growth picture of Marconi et al. (2004), and c) a non-evolution scenario for comparison.
The evolution of the accretion rate of model (b) has been derived by using the average BH growth history as published in Marconi et al. (2004) from redshift $z=3$ (where $100\%$ source activity has been assumed as an initial condition) to $z=0$ together with their supported accretion radiation efficiency of 0.1.
Model (a) and (b) both show strongly redshift-dependent BH accretion rates with higher rates at larger redshifts for 
BH masses of $10^8-10^9$M$_\sun$, typical for quasars.
For the present work the chosen models are used plainly as means and agencies to demonstrate how any evolution of 
accretion rates transforms into a redshift-dependence of the local optical pair production depth in strong-line quasars.

\section{Is the local opacity redshift-dependent?}

When studying the evolution of the EBL by means of a statistical analysis of signatures of $\gamma$-ray attenuation with effective opacity 
$\tau_{\rm eff}=\tau_{\rm EBL}+\tau_{\rm source}$ from extragalactic objects, recognizing and disentangling absorption taking place within the source system (''local opacity''; $\tau_{\rm source}$) and in the EBL during photon propagation to Earth (''EBL-caused opacity''; 
$\tau_{\rm EBL}$) is a crucial task. If both opacities each
depend on redshift in the same direction, but with the exact value of each being unknown, the redshift parameter alone will 
not be sufficient to extract the evolution of the EBL-caused opacity.
In order to assess the probability of this to take place in quasar-like AGN,
in the following I will apply realistic evolving, as well as non-evolving, cosmic accretion rate curves (see Fig.~\ref{fig2}) to the calculation of the expected opacity from $\gamma$-ray absorption in their accretion disk and BLR radiation fields, using the procedure outlined in Sect.~2. Any evolution of accretion rates translates into an evolution of the target photon fields
under consideration for photon-photon interactions, and thus into a redshift-dependence of the local opacity.
For this study the position of the $\gamma$-ray production is fixed to $l_0=0.01$pc. A location of the $\gamma$-ray emission region close to the BLR is particularly preferred by leptonic as well as hadronic blazar emission models that require external photon fields like accretion disk and BLR radiation as an ingredient for $\gamma$-ray production through particle-photon interactions. If 
$\gamma$-ray production is located well beyond the BLR, $\gamma\gamma$-pair production on external photon fields ceases to be an 
important process, and so will external Compton scattering above pair production threshold owing to their comparable cross section values there, or photopion production on accretion disk and BLR photons.

The goal of this excercise is to verify a possible {\it redshift-dependence} of the local opacity. The absolute $\tau_{\gamma\gamma}$ values depend on the details of the target radiation fields and location of the $\gamma$-ray production \citep[see e.g.][]{Donea03}, and thus come with uncertainties that reflect the dimension of the free parameter space.

\begin{figure}[t]
\resizebox{\hsize}{!}{\includegraphics{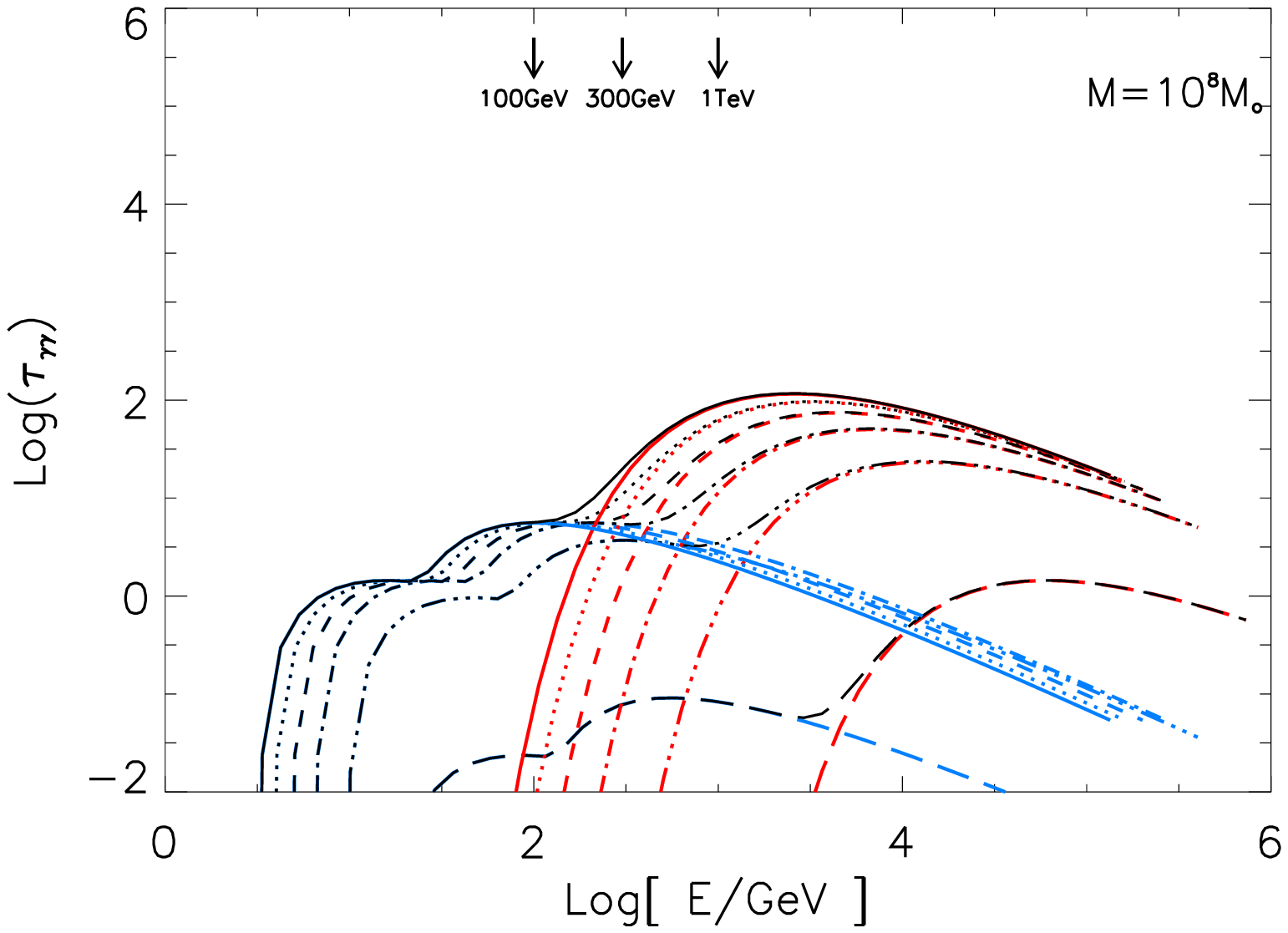}}
\caption{Optical depths from $l=0.01$pc to $l\rightarrow\infty$ for jet $\gamma$-rays interacting with accretion disk photons (red/pale grey curves) and BLR photons (blue/dark grey curves) with accretion rates following the Netzer \& Trakhtenbrot (2007) -- Lapi et al. (2006) evolution curves for
a M$_{\rm BH}=10^8$M$_\sun$ mass BH. The black curves represent the sum of both, opacity in the accretion disk and BLR radiation field, and correspond to redshift $z=0.1, 1, 2, 3, 4, 5$ from bottom to top, respectively. The fast evolution of accretion rates at low redshifts is apparent. Note that for a large energy range $\gamma$-ray absorption occurs mostly near the increasing part of the $\tau_{\gamma\gamma}(E)$ function near pair production threshold.}
\label{fig3}
\end{figure}

Fig.~\ref{fig3} shows the resulting opacity curves in the observer frame when using the evolutionary behaviour of accretion rates following
 Netzer \& Trakhtenbrot (2007) and Lapi et al. (2006) for the target photon fields for $\gamma\gamma$-interactions. 
The fast evolution of accretion rates at low redshifts is apparent. The corresponding curves for a 
M$_{\rm BH}=10^9$M$_\sun$ are inflated towards higher $\tau_{\gamma\gamma}(E)$ by more than an order of magnitude, and
for $E\geq 100$~GeV the photon-photon collisions occur predominantly in the accretion disk radiation field. 
Energy redshifting is the reason for the threshold and energy of maximum interaction probability,
$E^*\approx 1.2/(1+z) (\epsilon/{\rm eV})^{-1}$TeV, to decrease with redshift. 
Note that for most energies $<1$TeV $\gamma$-ray absorption occurs preferentially near the increasing part of the $\tau_{\gamma\gamma}(E)$ 
function near pair production threshold.
If the distance $l_0$ of the $\gamma$-ray production site from the disk rises beyond the BLR, $\gamma$-ray attenuation tails off
as indicated in Fig.\ref{fig1}.
An increase in BLR size leads to an opacity decrease for an unchanged BLR luminosity by an amount that
corresponds to the decrease in target photon density with increasing BLR volume.

\begin{figure}[t]
\resizebox{\hsize}{!}{\includegraphics{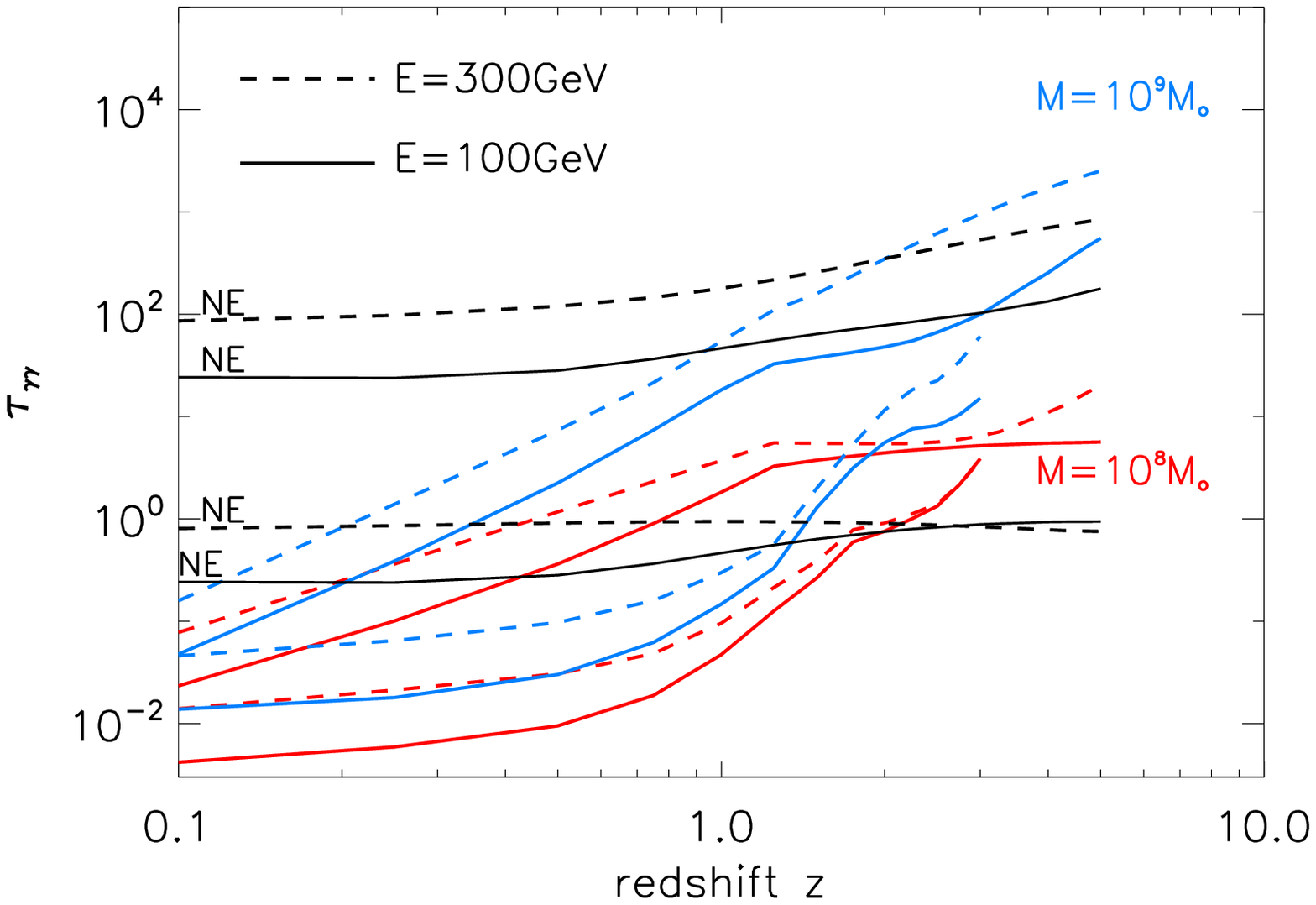}}
\caption{Redshift-dependence of ''local opacity'' of jet $\gamma$-rays interacting with accretion disk and BLR photons with accretion rates following the Netzer \& Trakhtenbrot (2007) -- Lapi et al. (2006) (upper colored curves; grey in printed paper version) and the Marconi et al. (2004) evolution model (lower colored curves; grey in printed paper version) compared to resulting opacities for non-evolving accretion rates (upper black curves 'NE': 
M$_{\rm BH}=10^9$M$_\sun$, $\dot M_{\rm acc}=M_{\rm edd}$, $L_{\rm disk}\approx 0.5L_{\rm edd}\approx 6\cdot 10^{46}$erg s$^{-1}$, lower black curves 'NE': M$_{\rm BH}=10^8$M$_\sun$, $\dot M_{\rm acc}=0.1M_{\rm edd}$, $L_{\rm disk}\approx 0.05L_{\rm edd}\approx 6\cdot 10^{44}$erg s$^{-1}$). A strong redshift-dependence is apparent in almost all cases, including the case of non-evolving accretion rates. The
rather constant $\tau_{\gamma\gamma}(z)$ curve at 300~GeV for the non-evolving low-rate low BH mass case is due to the $\gamma$-$\gamma$ 
interactions occuring predominantly in the valley between the $\tau_{\gamma\gamma}(E)$ curves for the BLR and accretion disk photon field 
(see Fig.~\ref{fig3}).
}
\label{fig4}
\end{figure}

A direct view on the redshift-dependence of the local opacity opens up by slicing Fig.~\ref{fig3} at the energies of interest. 
Fig.~\ref{fig4} shows the resulting $\tau_{\gamma\gamma}(z)$ curves at observer energy 100 GeV (all solid lines) and 300 GeV 
(all dashed lines) for all three evolution pictures for
the accretion rate as shown in Fig.~\ref{fig2} and for typical quasar BH masses (M$_{\rm BH}=10^8$M$_\sun$: all red curves, 
M$_{\rm BH}=10^9$M$_\sun$: all blue curves). The black curves represent the optical depths for the situation of a non-evolving 
high-accretion rate disk with a high-mass BH, and of a non-evolving low-accretion rate system with a lower-mass BH.
A strong redshift-dependence is apparent in almost all cases. Even for the case of non-evolving accretion rates the 
''local opacities'' show redshift-dependence, if $\gamma\gamma$ pair production occurs predominantly close to 
threshold, with an increasing slope of $\tau_{\gamma\gamma}(z)$ with redshift. The reason lies basically in the prominent
peak of the pair production cross section, together with cosmological energy red-shifting:
The prominent peak in the cross section leads to most photon-photon collisions occuring in a rather narrow
energy range $\Delta\epsilon^*$ near threshold for smooth broadband target photon spectra. If those source spectra are located at 
cosmological distances, the energy red-shifting into the observer frame leads to the presented redshift-dependence of 
$\tau_{\gamma\gamma}$.

\begin{figure}[t]
\resizebox{\hsize}{!}{\includegraphics{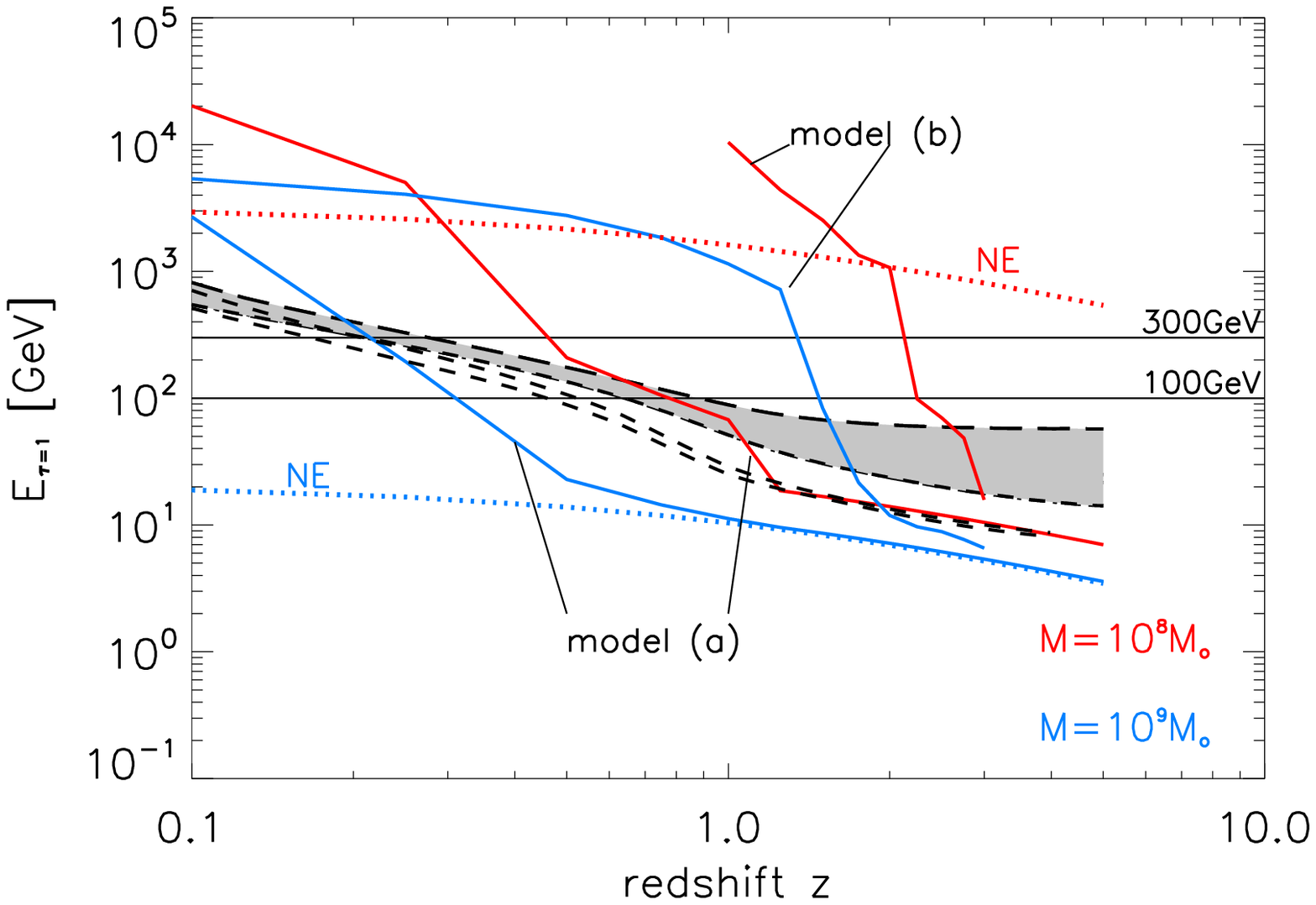}}
\caption{Critical energy $E(\tau_{\gamma\gamma}=1,z)$ versus redshift ($\gamma$-ray horizon) of jet $\gamma$-rays interacting with 
accretion disk and BLR photons with accretion rates following the Netzer \& Trakhtenbrot (2007) -- Lapi et al. (2006) (lower solid 
 colored curves; model (a)) 
and the Marconi et al. (2004) evolution picture (upper solid colored curves; model (b)), and for non-evolving accretion rates 
(dotted curves 'NE'
 for $\dot M_{\rm acc}=0.1$ (red/pale grey) and 1 (blue/dark grey)), compared to the Fazio-Stecker presentation of various evolutionary EBL models
 of Kneiske et al. (2004) (grey shaded area) and Stecker et al. (2006) (fast evolution model: upper short-dashed line; baseline model: lower short-dashed line). Exemplary observer energies of 100~GeV and 300~GeV are indicated by black solid lines.
 All models show a redshift-dependence of $E(\tau_{\gamma\gamma}=1)$. }
\label{fig5}
\end{figure}

It is straight forward to then determine the evolution of the e-folding cutoff energy $E(\tau_{\gamma\gamma}=1)(z)$ for the series
of accretion rate evolution models used so far.
Fig.~\ref{fig5} shows a Fazio-Stecker-like presentation for the local absorption and compares with the typically expected
behaviour for absorption of $\gamma$-rays in the evolving EBL \cite[e.g.][]{Primack99,Kneiske04,Stecker06}. Potential EBL probes of GLAST's LAT are located at a 
redshift $z\geq 0.5$. In all cases, and this includes also non-evolving
accretion rates, decreases the e-folding cutoff energy with redshift, similar to the Fazio-Stecker relation for
EBL-caused absorption. If local absorption
in the external radiation fields of AGN leaves measurable imprints in the $\gamma$-ray spectra, these will be almost unavoidably 
redshift-dependent, and remain to be distinguished from the $\gamma$-ray opacity of the EBL.


\section{Conclusions and discussion}

This work is devoted to investigate whether any redshift-dependence of the opacity $\tau_{\gamma\gamma}$ of $\gamma$-rays produced 
in systems of strong-line blazars can be uniquely attributed to photon absorption in the EBL. 
The possible existence of further sources of 
redshift-dependence of $\tau_{\gamma\gamma}$ other than caused during propagation in the EBL will lead to ambiguities in estimating
the EBL evolution.
Strong-line quasars are in general considered as high-luminosity sources in the $\gamma$-ray domain, which have a marked probability
of providing sufficient photon statistics to search for absorption breaks in a spectral analysis. 
Their bright external photon fields (i.e. accretion disk, BLR) are used in many high-energy emission models as a necessary 
ingredient for $\gamma$-ray production, e.g. via inverse Compton scattering. If this, however, takes place in the Klein-Nishina regime, and therefore above the pair production
threshold, $\gamma$-ray absorption due to photon-photon interactions is of comparable importance.
In some hadronic blazar models the observed $\gamma$-ray output is the result of re-distributing
the injected nucleon energy via pair cascades that develop in external photon fields, and thus
inevitably involves photon-photon pair production there.
The focus of this study is therefore directed to $\gamma\gamma$-pair production in photon environments
(here specifically accretion disk and BLR radiation field) of $\gamma$-ray loud blazars 
where the interactions of relativistic particles in external photon fields potentially contributes significantly to the $\gamma$-ray output above $\sim 50-100$~GeV.

The search for any redshift-dependence of this ''local'' $\gamma$-ray opacity leads to the following conclusions:
\begin{itemize}
\item If the $\gamma$-ray emission region is located not well beyond the BLR, mandatory for $\gamma$-ray production that involve external 
photon fields, local $\gamma$-ray absorption features in strong-line quasar spectra have to be expected at $E(1+z) >$ several tens 
of GeV. Local $\gamma$-ray absorption in external AGN radiation fields ceases importance if the $\gamma$-ray production site is sufficiently
distant from the BLR. While then evolutionary studies of the EBL by means of $\gamma$-ray absorption signatures are not affected, it will
lead to important implications for the high-energy blazar emission models.

\item Following recent progress in BH demographics, BH growth and corresponding accretion rates turn out to show a 
redshift-dependence
with higher rates at larger redshifts. Correspondingly, the critical energy $E(\tau_{\gamma\gamma}=1)$ due to local absorption in 
quasar disk and BLR radiation fields decreases with redshift, similar to Fazio-Stecker's relation for EBL absorption.

\item Even for the case of no evolution of quasar disk accretion rates, $E(\tau_{\gamma\gamma}=1)$ decreases with redshift for sources of a 
given BH mass, at $z>1$ very similar to the Fazio-Stecker presentation of EBL absorption with the current knowledge
of those systems. It results from the interplay 
of local absorption near pair production threshold and cosmological energy red-shifting.

\item Any observed redshift-dependence of absorption features in blazars, that are prone to local $\gamma$-ray absorption, can therefore
 {\it not} serve as a unique signature for absorption occurring in the EBL radiation field.
This complicates approaches for estimating the evolution of the EBL using GeV-sensitive instruments, that utilize the Fazio-Stecker 
relation or similar methods, and $\gamma$-ray AGN whose external photon fields are considered important in $\gamma$-ray production.
As a result, it seems that
only ''naked $\gamma$-ray jet sources'' (i.e. AGN without noticable optical/UV radiation fields close to the $\gamma$-ray emission region; 
``true type-2 AGN'') 
are unbiased probes for studies of the evolution of the EBL on the basis of the Fazio-Stecker relation and using GeV instruments like GLAST, etc.

\end{itemize}

Consequently, an obvious choice for suitable candidate sources for this task would be blazars with particular weak or absent
emission lines, generally classified as BL Lac objects (although exceptions exist, see Sect.~1). Predictions for the
expected number of GeV BL Lacs range from several hundred \cite{Dermer07} to a few thousand \cite{Muecke00b} above the
LAT sensitivity. It remains to be seen whether the near future GeV instruments will be detecting a sufficient number of
suitable sources at $z\geq 0.5$ to allow a sensible analysis on the basis of naked $\gamma$-ray jet sources only.

Though the finding of this work adds fundamentally to already recognized complications (e.g. flux and spectral source variability) in
analysis aiming to probe EBL evolution through the $\gamma$-ray horizon, it also offers new options for constraining 
AGN physics. The relevance of $\gamma$-ray absorption for cutting off the SED
at the high energy end in the different blazar types, possibly as a consequence of the location of
$\gamma$-ray production, could be probed by means of a statistical study of 
$\tau_{\gamma\gamma}(E,z)$ as a function of source type or object parameters. If simultaneously measured emission lines 
and/or accretion disk signatures indicate the presence of luminous photon fields external to the jet, the position of the $\gamma$-ray production site could be constrained. 
Monitoring both, the time history of the external target photon fields in AGN and the jet $\gamma$-ray flux, may offer independent verification of the importance of local absorption and extern inverse Compton scattering there, and allows
conclusions on the $\gamma$-ray production location and on some properties of the BLR material \cite{Boettcher95}.
At the same time, any non-detection of
absorption features with sensitive GeV instruments puts limits on the external radiation fields in those AGN.

If local opacity shapes part of the $\gamma$-ray loud AGN population, its evolution may influence luminosity function and 
extragalactic $\gamma$-ray background contribution of AGN above the 50-100~GeV energy range. In both cases a 
redshift-dependence of the local 
opacity of $\gamma$-ray loud quasars will have far-reaching implications.

\acknowledgments

{\bf Acknowledgments}

I'd like to thank for valuable feedback
from the GLAST-LAT AGN science working group, in particular for very useful comments from C. Dermer, G. Madejski and B. Lott.
I also thank F. Stecker for providing his model curves and for interesting discussions.
 This work is supported by the National Aeronautics and Space Administration under contract NAS5-00147 with Stanford University.



\end{document}